# Superconducting Cables Characterization with an electrical method

L. Morici[1], E. Tamburo De Bella[2] and G. Messina[1]

## ABSTRACT

AC losses in High Temperature Superconductor tapes (HTS) are an important design parameter for large scale power applications. The electrical method is often used for losses measurements on SC tapes and, when the results are to be compared to computed estimations, simple theoretical models like the Norris-Brandt are typically used. Electromagnetic dynamic effects aren't usually taken into account by the simpler models, but their presence may corrupt the measured data. In this paper, a preliminary study on a conventional copper bar sample has been done to evaluate the effect of the skin currents on the specimen losses measures obtained from the electrical method. The measurement procedure has afterward been applied to a SC YBCO tape (Superpower® SCS4050AP), obtaining circuit configuration dependent results. Because the SC losses must obviously be measurement circuit independent, a systematic error due to the skin currents must be playing a role. The skin current induced error is always there and, in order to apply the electrical method for losses estimation, the only possibility is its minimization. The best choice for the measurement circuit to be adopted, in order to minimize the skin current undesired contribution, will finally be proposed.

**Index terms:**
ac losses, AC (alternating current), DC (continuous current), SC (superconductors), numerical modeling, analytic modeling, hysteresis losses, coupling losses, magnetic materials, 2D (two dimensions), 2G ($2^{nd}$ generation, 3G $3^{rd}$ generation, ...), rms (root mean square), LN (liquid nitrogen), S/N (signal to noise ratio), T (temperature), Vibrating Sample Magnetometer (VSM)

## INTRODUCTION

For a proper design of power applications based on High Temperature Superconductors (HTS) coated-conductor tapes, AC losses calculation and measurements is mandatory. The term "AC losses" meaning "the energy dissipated per cycle and per unit length of the superconducting tape". AC losses mechanisms of the SC layer inside a coated conductor tape are typically hysteretic, but the SC tape is a complex system, made up of various layers of different materials each one responsible of some sort of energy losses as a response to external stresses, and various contributions to the overall losses are expected (eddy current on metal layers, ferromagnetic losses on the magnetic substrate, etc.). In any case, the SC layer hysteretic losses are expected to prevail in the frequency range explored here and, as a consequence, will be the only computed losses contribution directly compared to the measures.

Electromagnetic and calorimetric methods are typically used for AC losses measurements. In particular, the HTS sample is exposed to an alternating magnetic field in the VSM experimental setup, or is subjected to transport current in the electrical method experimental setup, while the calorimetric method sensibility is too poor for single tape characterizations (the calorimetric method is usually applied to HTS component losses measures: coils, windings, etc.).

Here we'll report on the implementation, realized c/o the ENEA laboratories in Frascati, of an experimental apparatus for the electrical method characterization of bare SC tapes. For comparison purposes, a copper bar has been used as a reference sample. Once validated with the copper bar, the measurement system has been used for the superconducting tape characterization.

---

[1] L. Morici and G. Messina are with ENEA EUROfusion, via Enrico Fermi 45, 00044 Frascati, Roma, Italy
E. Tamburo De Bella is with Department of electrical Engineering and Information Technology, University Tor Vergata, Roma, Italy

Finally, the obtained losses have been compared to the theoretical predictions of the hysteretic Norris-Brandt model [1].

It is crucial, for the successful application of the electrical method, to have a Reference signal in-phase with the sample current. The measured signal, extracted from voltage taps positioned on the sample, will then be subdivided among the Reference in-phase (Real) and the Reference out of phase (Imaginary) component: the in-phase component has been used for losses estimation while the out of phase signal, mostly inductive (and, as a consequence, measurement circuit configuration dependent), may be used for inductance estimations.

To investigate the behavior of the transport current in-phase voltage component of the measured signal, various paralleled acquisition wires have been positioned beside and on each sample, and the measuring signals coming from the acquisition wires have been compared among themselves.

Due to the systematic effect induced from a component of the skin currents on the pickup wires, circuit configuration dependent losses results were expected from the copper bar (even if the true losses must obviously be measurement circuit independent). The AC losses measured for the superconducting tape must be measurement circuit independent as well but, similarly to the copper bar, spurious voltage signals of inductive origin (in-phase with the current) due to the skin currents may arise. Part of the skin signal is indistinguishable from the voltages due to the losses, and is responsible for wrong losses estimations.

Skin signals (originating measurement errors) are always present, but a proper choice of the measurement circuit configuration may minimize the skin contribution. The best measurement circuit configuration to be used will finally be proposed and, as a further check of the reliability of the obtained measures, the measurement results have been compared to the computation coming from the Norris-Brandt theoretical model.

## EXPERIMENTAL SETUP

### *ELECTRIC CIRCUIT DESCRIPTION*

The implemented experimental apparatus, developed c/o the Frascati ENEA labs., can be divided in two sections: Power and Acquisition (Fig.1).

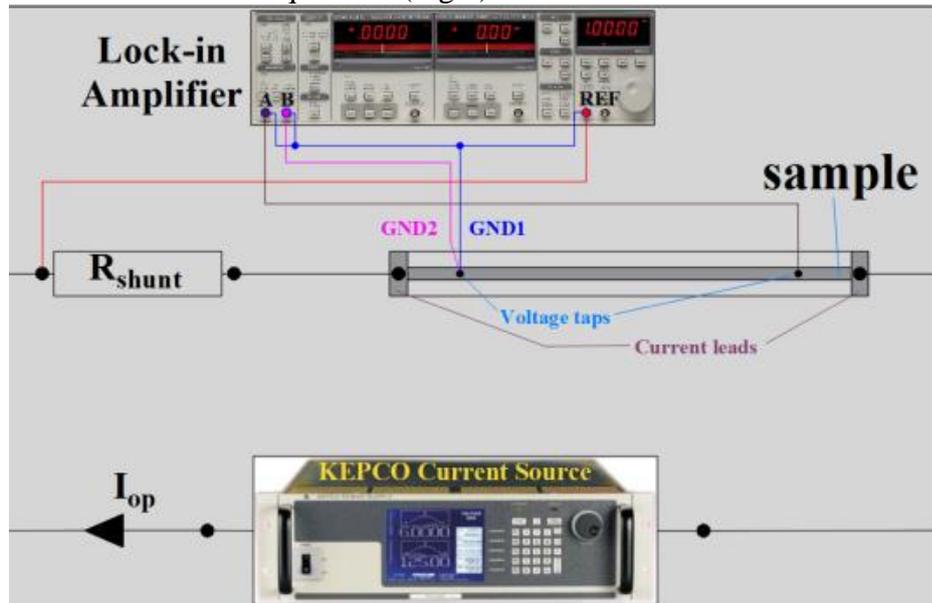

**Fig. 1: schematic of the used measurement apparatus**

In the power section, a KEPCO programmable amplitude (*0 ÷ 50 Apeak*) and frequency (*0 ÷ 400 Hz*) current source is used to supply current, $I_{op}$, to a shunt resistor connected in series to a sample (the used samples are an HTS tape or a copper bar). The shunt resistor ($R_{SHUNT} = 200\ m\Omega$) voltage has been used for accurate current estimation and, due to its temperature dependence, a calibration curve has been obtained in advance: $R_{SHUNT} = R_0 + \gamma T$ (where, for the used apparatus: $\gamma = 2 \cdot 10^{-4} [\Omega/°C]$, $R_0 = 0,2 [\Omega]$. The Temperature (*T*) was obviously recorded at each measure and used for $R_{SHUNT}$ correction estimation. The parasitic parameters of the components used in the circuits, if not considered, may be sources of systematic measurement errors. Among them, the series inductance of $R_{SHUNT}$, namely $L_{SHUNT}$ (not shown in Fig.1), is of particular importance and has been measured using an Agilent E4980A LRC meter (and the result used for compensation purposes during data analysis). In the analyzed frequency range (*0÷400 Hz*), the measured system equivalent series inductance of $R_{SHUNT}$ resulted to be: $L_{SHUNT} = 3,5 \cdot 10^{-6}$ H.

Two GND wires, welded on the sample at the same tap point and immediately twisted afterward, named $GND_1$ and $GND_2$, have been used as reference points. The first, arbitrarily called $GND_1$, used as the chassis voltage for the instrumentation (i.e. the lock-in SR830); the second, $GND_2$, used as the input signal on one of the two channels of the lock-in amplifier. The Stanford SR830 dual channel lock-in inputs have been set in floating mode, and the differential voltage across the two voltage taps on the sample has been measured.

About the AC losses measurement methodology, the usual electrical procedure was adopted: the voltage across $R_{SHUNT}$ (once corrected for parasitic contributions and the temperature) is assumed to be aligned to the operative current, *Iop*, and has been used as the lock-in Reference signal; the circuit in-phase and out of phase signals, referred to *Iop* and coming from the voltage taps on the sample, are the quantities of interest.

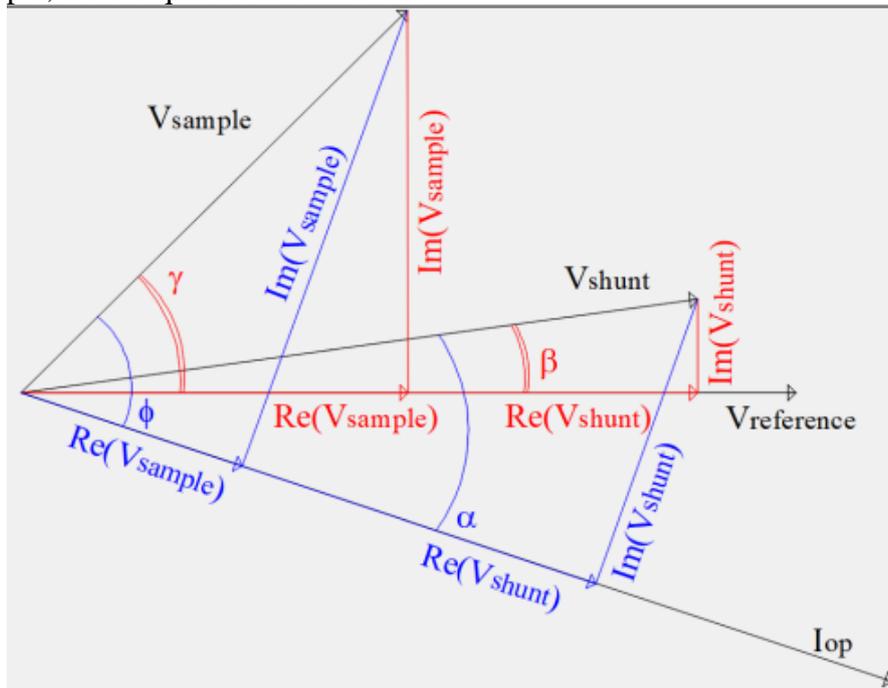

**Fig.2: inserting the same signal on the lock-in input channels (V<sub>shunt</sub>) and reference channel (V<sub>reference</sub>), a fictitious quadrature signal to be compensated (defining Re(V<sub>shunt</sub>) and Im(V<sub>shunt</sub>) i.e. the angle β) is obtained. Now, at this stage, the V<sub>sample</sub> signal coming from the SC is referred to V<sub>reference</sub> and the angle γ is defined. But everything must be referred to the operating current, *I*<sub>op</sub>, and, to this aim, another compensation angle (given by $\alpha = \omega L_{SHUNT}/R_{SHUNT}$) due to the parasitic inductance of the power circuit, must be used. Finally, from the measured quantities of V<sub>sample</sub>, its modulus can be extracted and, with the angle $\phi = \gamma - \beta + \alpha$ the Re(V<sub>sample</sub>) component to be used for losses computation is obtained. A similar computation is applied to obtain I<sub>op</sub> from Re(V<sub>shunt</sub>) and Im(V<sub>shunt</sub>).**

With reference to the purely qualitative Fig.2 vector diagram, both the in-phase red Re(V<sub>sample</sub>) signals (used for losses estimation), and the out of phase red Im(V<sub>sample</sub>) signals (used for sample inductances estimation) have been compensated (blue lines); the same treatment has been applied to the $V_{shunt}$ components.

### *SAMPLES PREPARATION*

A fiberglass sample holder has been used to block the sample under measurement to the apparatus. The copper bar, having a square cross section *30 mm x 5 mm*, was used for system validation while the superconducting tape, (Superpower® YBCO-coated conductor commercial sample, SCS4050AP) was used for SC characterization. The main SC tape characteristics are: a strip (*1 μm* high, *4 mm* wide), a copper matrix stabilizer and a no-magnetic substrate (Hastelloy).

Both samples were about 1 m long, which is roughly the distance between the apparatus current leads. The voltage taps for signal detection have been placed at *60 cm* relative distance somewhere in between the current leads Fig.3. In this way, there are at least two order of magnitude between the relevant lengths of the superconducting sample ($10^{-6}$, $4 \cdot 10^{-3}$, $6 \cdot 10^{-1}$ m), and the 2D Norris-Brandt bare tape approximation [1] for the superconducting tape can be considered accurately fulfilled.

All the measures have been carried out at liquid nitrogen temperature and, to avoid magnetic couplings with the environment, a no-metallic cryostat has been used and any conducting or ferromagnetic material close to the sample has been removed. Particular care was aimed at the electrical reference point choice, conveniently defined to be one of the voltages tap measurement points on the samples, carefully avoiding any sort of ground loop in the assembly.

A description of the adopted measurement loop is also useful due to the importance given to the measurement circuit configuration in literature [4, 5]. As can be seen in Fig.3, the sample voltage signal was picked up using sequentially one among seven (for the superconductor), or one among six (for the copper bar), *0.25 mm* enameled copper wire welded at one voltage tap, aligned in parallel to the sample and among themselves up to the other tap *60 cm* far away, and positioned at a constant relative distance among each other while above the sample.

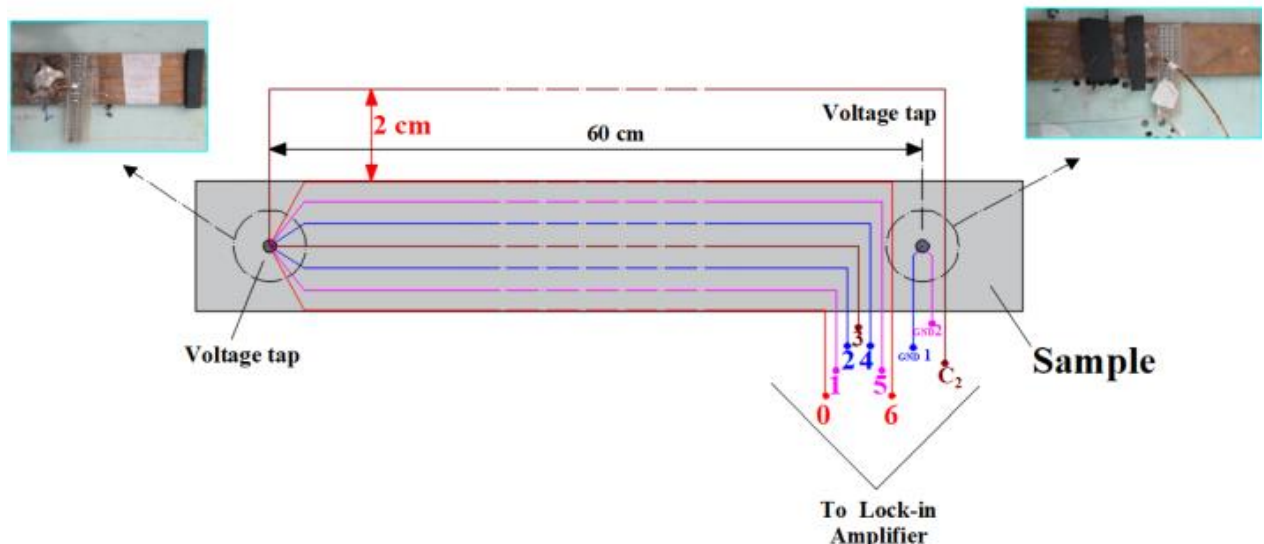

**Fig. 3: schematic of the used wiring configuration**

In addition, various external rectangular shaped turns welded at the same pick up point, placed at different distances from the sample up to the second voltage tap, were installed. The naming convention for these external pickup coils is *Cx*, (*x* is the distance, in cm, among the cable and the sample side). As mentioned above, the second voltage tap is *60 cm* far from the pick-up one and is chosen as the electrical reference point of the experimental setup. All the copper wires (namely, all the measurement circuits) have been glued to the sample holder or to the sample itself, using Araldite®. At one end they're welded to the sample, while at the other end they're twisted with the $GND_1$ and $GND_2$ wires and get ready for instrumentation connection. Of course, the validation and the measurement configuration circuits are very similar and conceptually identical.

## THEORETICAL ANALYSIS

A copper bar has been used for the measurement system validation. As is well known, the skin effect, as well as the ohmic contribution, usually affects the losses results of such a system. According to the skin effect theory [6], a good low frequency approximation for the current density inside a normal conductor bar of rectangular cross section ($a,b$) is given by the expression:

$$J_z(x,y) = \frac{I}{a \cdot b}\left(1 + j\omega\mu_0\gamma \frac{b}{a+b}\frac{x^2}{2}\right)\left(1 + j\omega\mu_0\gamma \frac{a}{a+b}\frac{y^2}{2}\right) \quad (1)$$

where $I$ is the total current on the bar, $a = 5\ mm$ and $b = 30\ mm$ are the sizes of the bar, $\omega$ is the angular frequency, $\mu_0$ the vacuum permeability, $\gamma$ the copper conductivity and the ranges of the position are $-a/2 \leq x \leq a/2, -b/2 \leq y \leq b/2$.

As can be seen from (1), the imaginary component of the skin current is zero on the bar axis, and reaches a maximum value on the external corners but, most importantly, such a component couples with the circuits giving rise to an induced signal in-phase with the operative current (i.e. the Reference signal). Circuit configuration losses results are then expected, due to this undesired inductive behavior, giving rise to a systematic error in the losses estimation when using the electrical method. Because such an error is of induction origin, it's also expected to be frequency dependent. Moreover, the reference signal coming from $R_{SHUNT}$ is "skin affected" as well. The systematic error coming from the skin effect cannot then be removed, and is responsible for the addition or subtraction of a contribution to the losses signal we're interested in, producing an underestimation or overestimation of the measured losses as a result; the only possibility to obtain meaningful data is the adoption of circuit configurations minimizing the skin contribution.

The device under examination is an SC tape, and expressions for AC losses on SC have been derived for a number of geometries, including the tape-like strip of our concern here. Experimental and theoretical approaches for losses estimation have been developed and, among the theoretical methods, it's worth mentioning the classical works by Norris [1] and Brandt [2]. The Norris-Brandt model takes into account a 2D superconducting layer indefinitely long, having a certain width and negligible height. Modern SC tapes, having millimetric width and micrometric height of the superconducting strip, adhere very well to the Norris-Brandt approximation, provided a straight length of the order of a meter is used in the experiments.

Among the approximations used by Norris-Brandt, the London hypothesis of a fixed (magnetic field independent) critical current looks rather rough, and a certain degree of disagreement between theory and experiment may be expected as a consequence. Another limitation of the Norris-Brandt model is its "quasi-static" nature, producing "in-phase" conduction current all over the cross section of the tape: all dynamic electromagnetic interactions, responsible for the skin effect on conventional conductors, aren't included in the model.

Even if more sophisticated approaches developed by the theoretical community (extensively described by Grilli & al. [3]) are available, and despite its limitations, the Norris-Brandt model is very useful for the intuition of the basic physics involved in the system, and its predictions will be used here for comparison with the data measured from the analyzed SC sample.

In the Norris-Brandt strip model, the tape loss per cycle per meter, $L_c$, is given by:

$$L_c = \frac{I_c^2 \mu_0}{\pi}\left\{(1-x)\ln(1-x) + (1+x)\ln(1+x) - x^2\right\} \quad (2)$$

$I_c$ is the tape critical current, $x = I_p/I_C$ the peak to critical current ratio, $\mu_0$ and $\pi$ the usual quantities. According to the theory, then, the only tape parameter determining the AC losses on a SC tape is the critical current $I_c$, the losses per cycle are then determined from the current peak $I_p$ to $I_c$ ratio. The SCS4050AP tape critical current at LN temperature is $I_c = 165\ A$, a value provided by the manufacturer and confirmed directly measuring the voltage across the whole tape length ($d = 60\ cm$) in DC conditions, using the 1 $\mu$Vcm$^{-1}$ threshold criteria for $I_c$ determination.

## EXPERIMENTAL RESULTS

### *MEASUREMENT SYSTEM VALIDATIONS*

In this preliminary step, a copper bar has been used to validate the experimental apparatus. It is well known that, in the low frequency range, the expected losses per cycle on the bar should mainly be due to the ohmic contribution, while the skin effect should prevail at higher frequencies. To study the contribution on the losses of the different components, an assembly similar to the one shown in Fig.3 has been adopted, namely: six wires, aligned among each other, plus one external wire, have been positioned over the bar and *2 cm* outside the bar, respectively. The sequence "0" to "5" has been used to identify each circuit on the bar and in particular wire "0" and wire "5" are aligned to the opposite external borders of the bar, while all the other wires are placed at a 5 mm distance among each other (wires "2" and "3" are the closest to the bar axis).

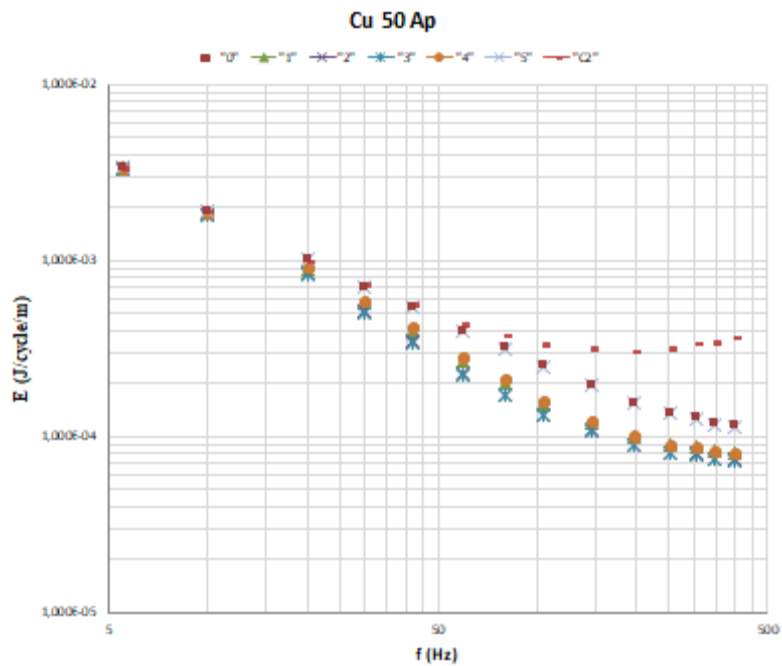

**Fig. 4: measured losses for the copper bar. The circuit configuration dependence with frequency is due to skin currents.**

The energy losses per cycle and per unit length obtained from the copper bar, using the current ($I_{op}$) in-phase rms component of V$_{sample}$, are shown in Fig.4. A measurement circuit configuration independent ohmic trend can be observed in the low frequency range, from which a value for the copper resistivity *($\rho = 1.35\ 10^{-9}\ \Omega m$)*, compatible with the expected value at liquid nitrogen temperature for the material, can be obtained. Conversely, the circuit configuration dependences of the curves in the high frequency range is the result of the skin effect.

Due to the symmetry properties of the magnetic field mapping on the bar surface, a reduction of the induced signal of skin origin is expected for the wires positioned over the bar. As a consequence, the signal coming from a circuit on the symmetric position, i.e. on the bar surface and aligned to its axis, gives the most reliable estimation of the losses measured with the electrical method, the skin contributions resulting minimized.

The measured out of phase signal of $V_{sample}$ can be ascribed to reactive contributions of inductive nature. It is well known that the reactive signal is measurement circuit configuration dependent and, at least for a conventional straight wire of cylindrical shape, a distinction can be made among internal and total inductance per meter. The expected DC internal inductance per meter of a cylindrical wire is also well known to be $L_{internal} = \mu_0/8\pi = 5\ 10^{-8}\ H/m$.

For conventional conductors of rectangular cross section, approximate solutions of the magnetic field distribution on the surface had only recently been published [6], the main problem coming from the boundary condition definition. The magnetic field distribution on the surface is strongly position dependent, and a single parameter is no longer sufficient for the out of phase signal characterization. In any case, due to the lack of a proper name, the term "internal inductance" to characterize the relationship between the out of phase measured voltage and the current will be used. In other words, if a conventional conductor of rectangular cross section were studied, an internal inductance dependent on the position of the measuring wire on the conductor surface as well as a frequency dependence, is to be expected. The "internal" inductance values, plotted in Fig.5, are immediately obtained from the relation:

$$L = \frac{\text{Im}(V_{tape})}{2\pi f \cdot I_{op} \cdot l} \qquad (3)$$

$Im(V_{tape})$ is the out of phase component of length normalized $V_{sample}$, , $f$ the signal frequency, $l$ the distance among voltage taps and $I_{op} = V_{shunt}/R_{shunt}$ the rms operating current. In the frequency range *(0 ÷ 400) Hz*, on the "low frequency" side of the spectrum, the estimated $L$ confirms the expected behavior. As can be seen in Fig.5:

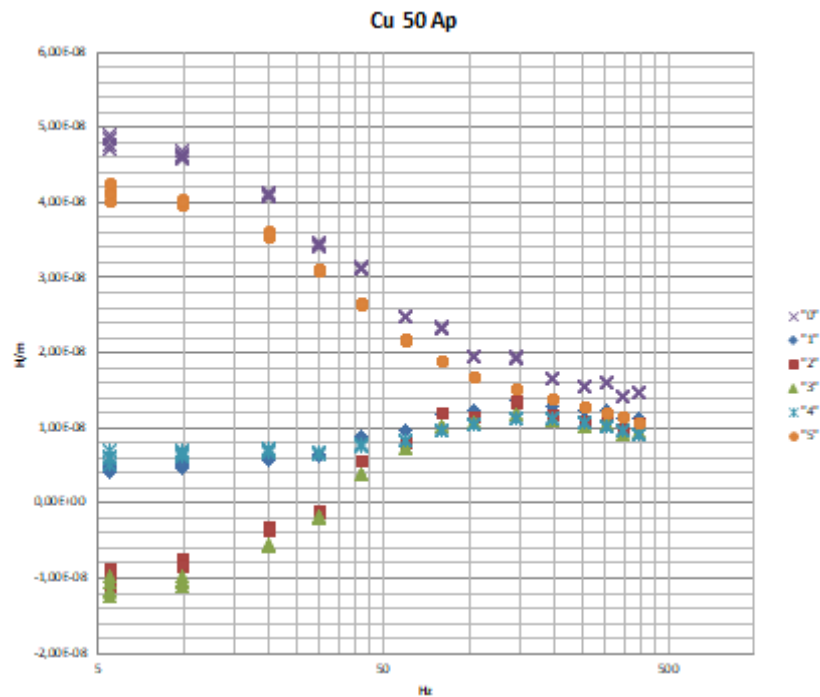

**Fig. 5: internal inductance values for the copper bar (plotted above are the results obtained for 50App current but, because of the current independence, the plot is representative of the results obtained from all the other currents)**

1. values on the expected order of $10^{-8} H/m$ for the internal inductance are obtained,

2. signals coming from symmetric circuits with respect the bar axis gives rise to similar internal inductances
3. not shown in Fig.5, the measures coming from the external circuit $C_2$ (the circuit *2 cm far apart the bar*) are one order of magnitude stronger than the plotted ones, as expected, but otherwise useless to our aims.

The obtained results from the simple copper bar are in agreement with the expected behavior, validating this way the utilized experimental apparatus.

## AC LOSSES ON A SUPERCONDUCTING TAPE

To characterize the SC, the lock-in measurement procedure has been applied to a sample at various frequencies and different currents. The power circuit has been driven setting up the amplitude and frequency of the KEPCO power supply to the desired current value, $I_{op}$. The real and imaginary components of the $V_{shunt}$, and of the $V_{sample}$, as measured from the lock-in, have been recorded.

As expected, the internal inductance results are current independent and circuit configuration dependent (Fig.6). Compared to the analogous results from the copper bar, a rather smooth decrease of the internal inductance values on the high frequency side of the excited spectrum may be observed.

AC transport current losses in the superconducting tape have been evaluated from the relation $E_{loss} = [Re(V_{sample}) \cdot I_{op}]/f \cdot l_{sample})$. The obtained results are plotted in Fig.7-10 for various driving currents. In the low current range (*10, 20 $A_{peak}$*), the KEPCO power supply unit generates a rather flat *rms* current signal all over the inspected frequency range. For higher currents (*40, 50 $A_{peak}$*), the KEPCO unit generates a current with an increasing *rms* trend at the higher frequencies: this behavior can be seen from the computed Norris-Brandt plotted lines in Fig.7-10, representing the theoretical values obtained inserting the measured current value in (2).

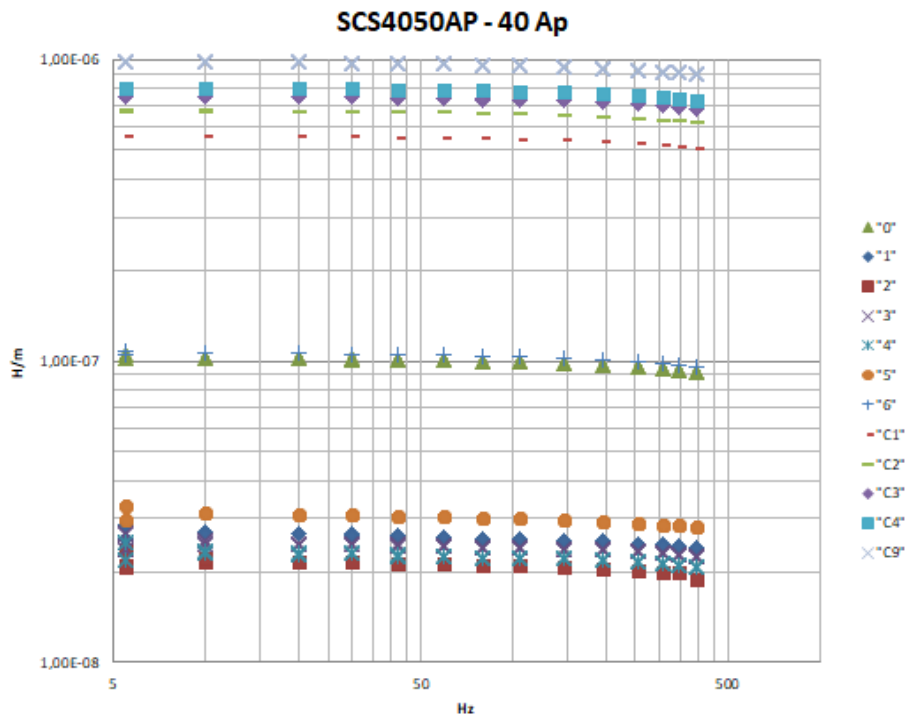

**Fig. 6: Internal inductance data for the SC sample. Only the measures for 40Ap have been shown. These results, current independent, are representative of all the driving currents.**

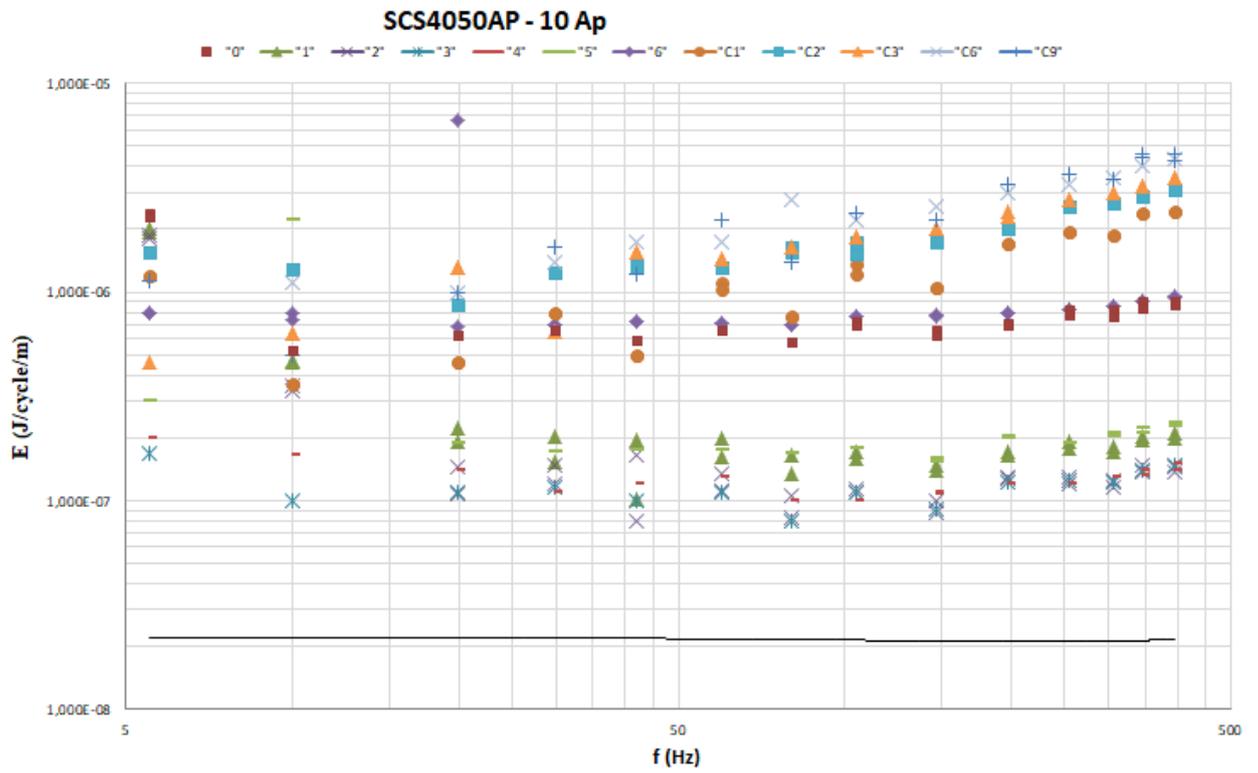

**Fig. 7:** measured losses with 10Apeak circuit current (the straight line is the Norris-Brandt computation).

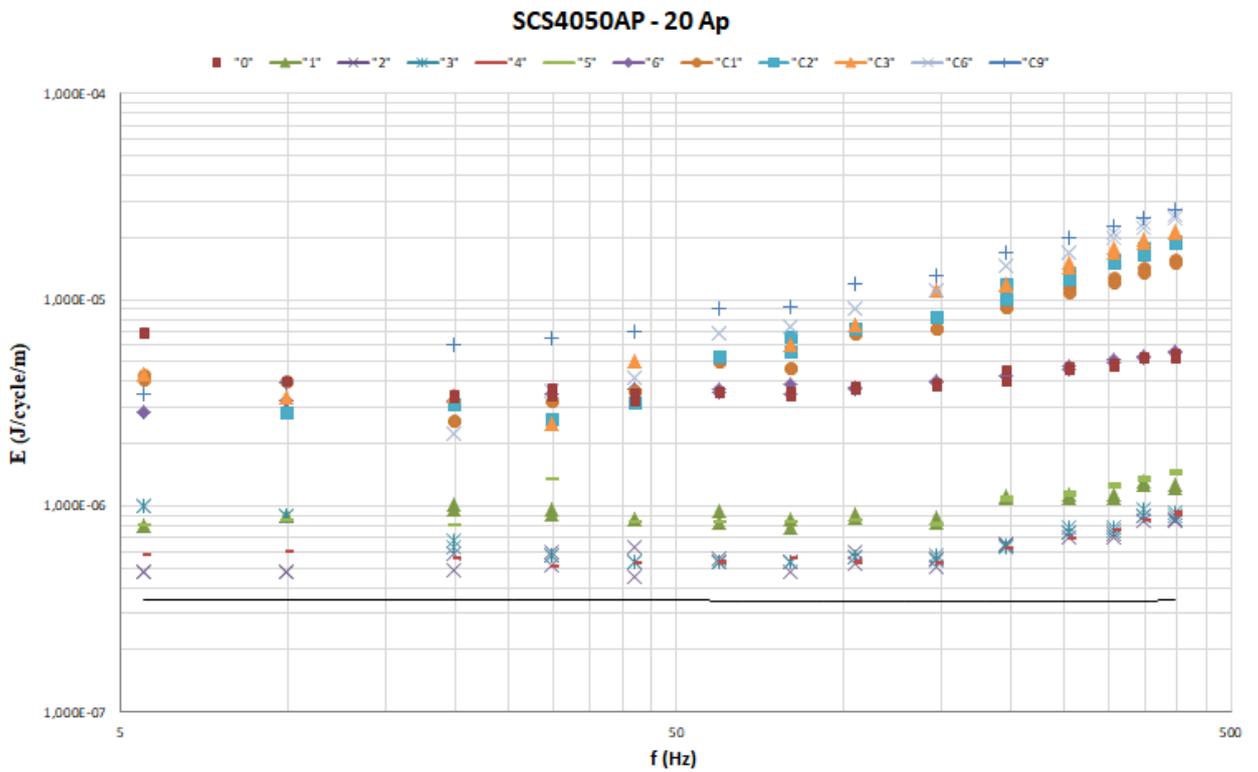

**Fig. 8:** measured losses with 20Apeak circuit current (the straight line is the Norris-Brandt computation).

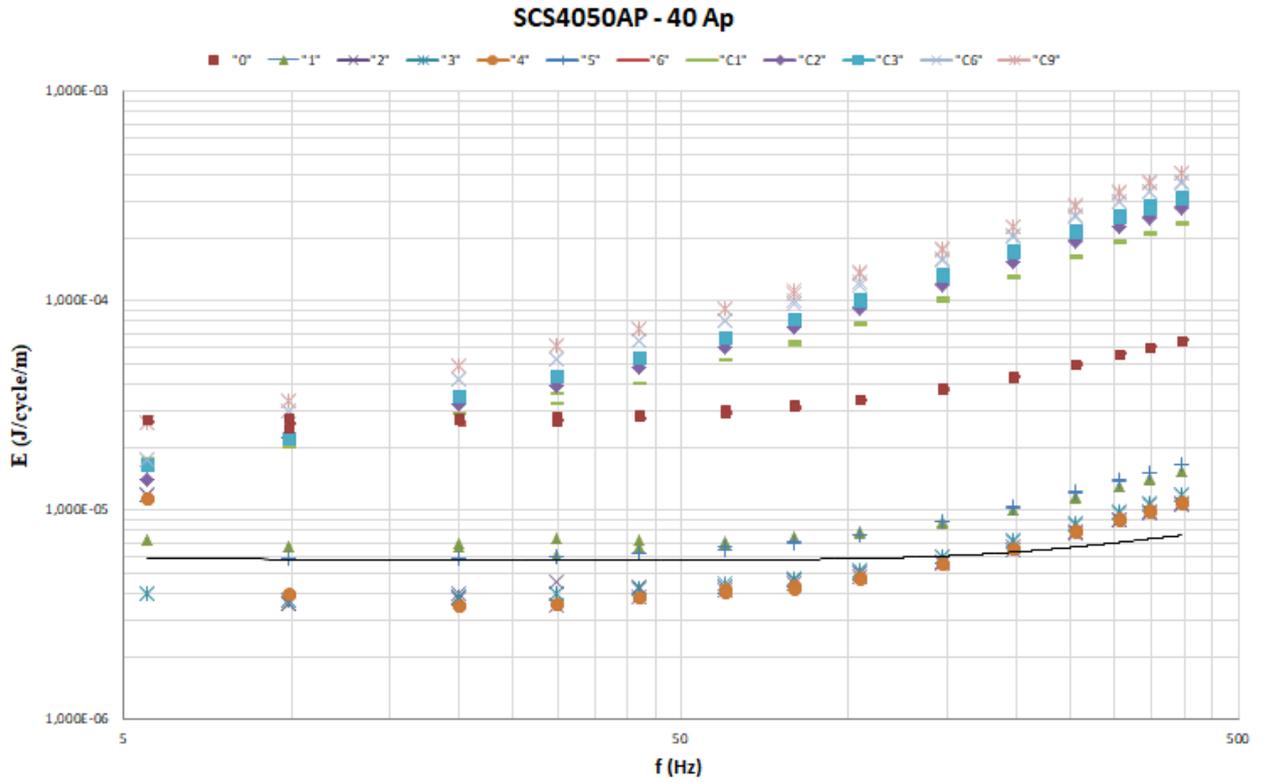

**Fig. 9: measured losses with 40Apeak circuit current (the straight line is the Norris-Brandt computation).**

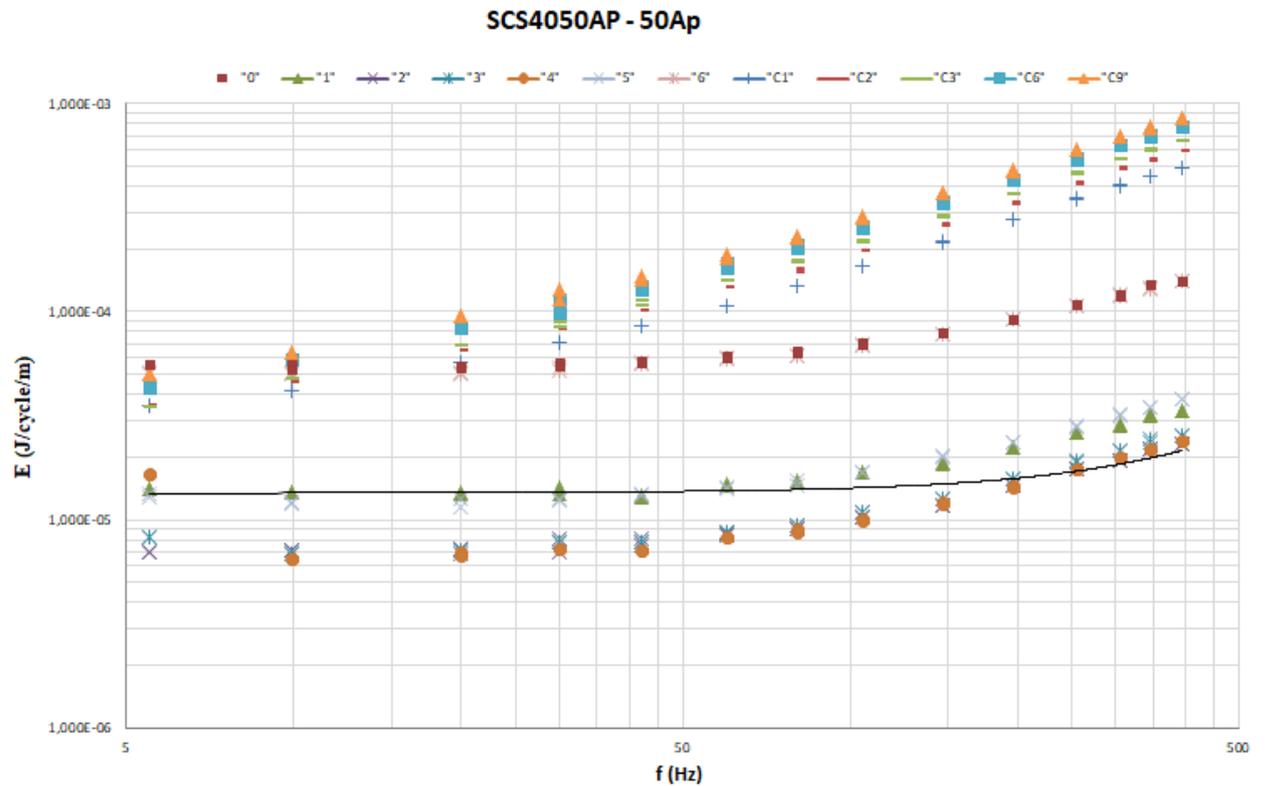

**Fig. 10: measured losses with 50Apeak circuit current (the straight line is the Norris-Brandt computation).**

All the measured losses, besides the circuit configuration dependence, also show a different behavior when compared with the Norris-Brandt computation. For low driving currents (namely, the *10 A$_{peak}$*), the measures are well above the theoretical expectation, a possible explanation could be the strong Bean [8] contribution necessary to generate the screening current nullifying the magnetic field inside the SC. A frequency dependence of the losses can be observed while increasing the driving current: the experimental data increasing with frequency (Fig.9 and Fig.10), a behavior probably due to skin currents.

Among the obtained losses values, those coming from the symmetric circuit positioned on the tape axis (namely, SC tape wire "3") give the closest indication of the tape losses. The symmetric circuit is then the best choice when the electrical method is used. Wire 3 results, for all the measured currents and frequencies, are plotted in Fig.11. As can be seen, the Norris-Brandt estimation is approached from the experimental results when increasing the supplied operative current, $I_{op}$. The accordance, always within an order of magnitude, can be considered rather good considering the approximations of the model.

Unfortunately, due to instrumentation limitations, the maximum applied driving current in our measures was *50 Apeak*, only 1/3 of the SCS4050AP tape critical current. Further investigations on different tapes and for higher currents will therefore be scheduled.

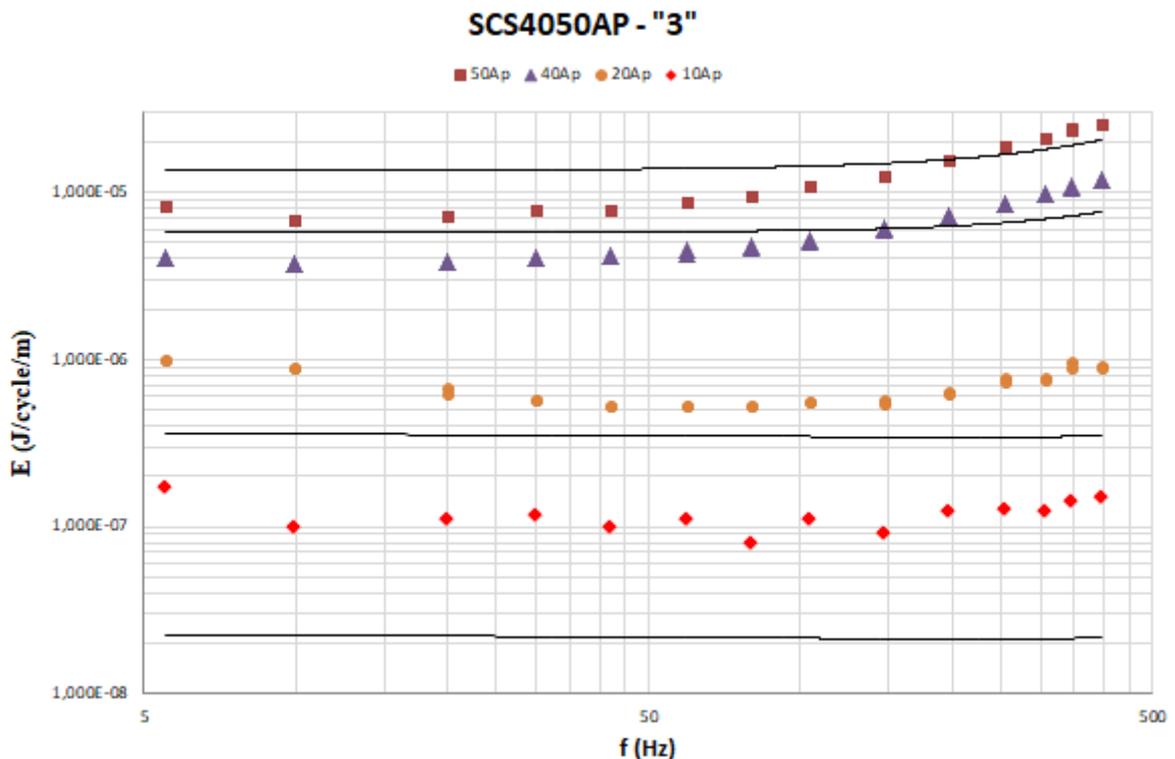

**Fig. 11. measured losses from the symmetric circuit (wire n."3") at various operating currents and compared to the theoretical prediction of the Norris-Brandt model.**

## CONCLUSION

A study has been reported on the losses measured on a bare SC tape using the electrical method. Measurement circuit configuration dependent results, likely due to skin currents, were obtained. Among the studied measurement circuits, the one minimizing the undesired skin contribution has been identified and proposed to be adopted.